\documentclass[twocolumn]{aastex701}
\usepackage{hyperref}
\usepackage{orcidlink}
\usepackage{amsmath}
\usepackage{comment}

\usepackage[utf8]{inputenc}
\usepackage{float}
\usepackage{natbib}

\begin{document}
\title{
 Rattle-and-Break: the Impact of Planetesimal Scattering on Super-Earth Resonant Chains
}
\author[0000-0002-1032-0783]{Sam Hadden}
\affiliation{Canadian Institute for Theoretical Astrophysics, 60 St George St Toronto, ON M5S 3H8, Canada}
\email{hadden@cita.utoronto.ca}
\author[0000-0003-0511-0893]{Yanqin Wu}
\affiliation{Department of Astronomy \& Astrophysics, University of 
Toronto, Toronto, Canada}
\email{wu@astro.utoronto.ca}
\begin{abstract}
The spacings of super-Earths in multi-transiting systems exhibit a distribution that is broad and mostly featureless, with the exception of notable excesses of planet pairs situated a few percent wide of first-order mean motion resonances (MMRs). In this work, we extend the so-called ``breaking-the-chains" model to account for both of these characteristics. Assuming that super-Earths are settled into stable chains of resonances after disk-driven migration, we show that scattering a planetesimal population that contains  only a few percent of a system's mass can reorganize  primordial chains in remarkable ways.  The planetesimal scattering ``rattles" the chains by  repelling adjacent planet pairs wide of their initial MMRs. Some chains remain rattled but otherwise intact and make up the observed excesses wide of MMRs. In other systems, however, this initial rattling sows the seeds of later orbital instabilities that break the chains entirely.  If individual planetesimals' masses are of order a Pluto mass or so,  the onset of these instabilities can occur tens or hundreds of Myr after birth, naturally explaining the apparent disappearance of near-resonant pairs on this timescale. The origin of such Pluto-mass debris is currently unknown.
\end{abstract}

\section{Introduction}
\label{sec:intro} 

Systems of super-Earths  at close orbital distances ($a\sim 0.1-1\,\mathrm{AU}$) from their host star are common around sun-like stars
\citep[e.g.,][]{Fressin2013,Petigura2018,Zhu2018}. How such planets form, and why our own solar system lacks any such planets, remain open questions.  One of the central  questions  is how their current orbital architectures arise. 

 Currently, there are two main puzzles regarding super-Earth system architectures: the lack (but not absence) of exact mean-motion resonance pairs and the apparent pile-ups just wide of these resonances. We review these puzzles briefly, together with previous proposals to explain them, before introducing our own idea.  Our work here aims to explain these two puzzles in one framework.

 \begin{figure*}
    \centering
    \includegraphics[width=0.9\linewidth]{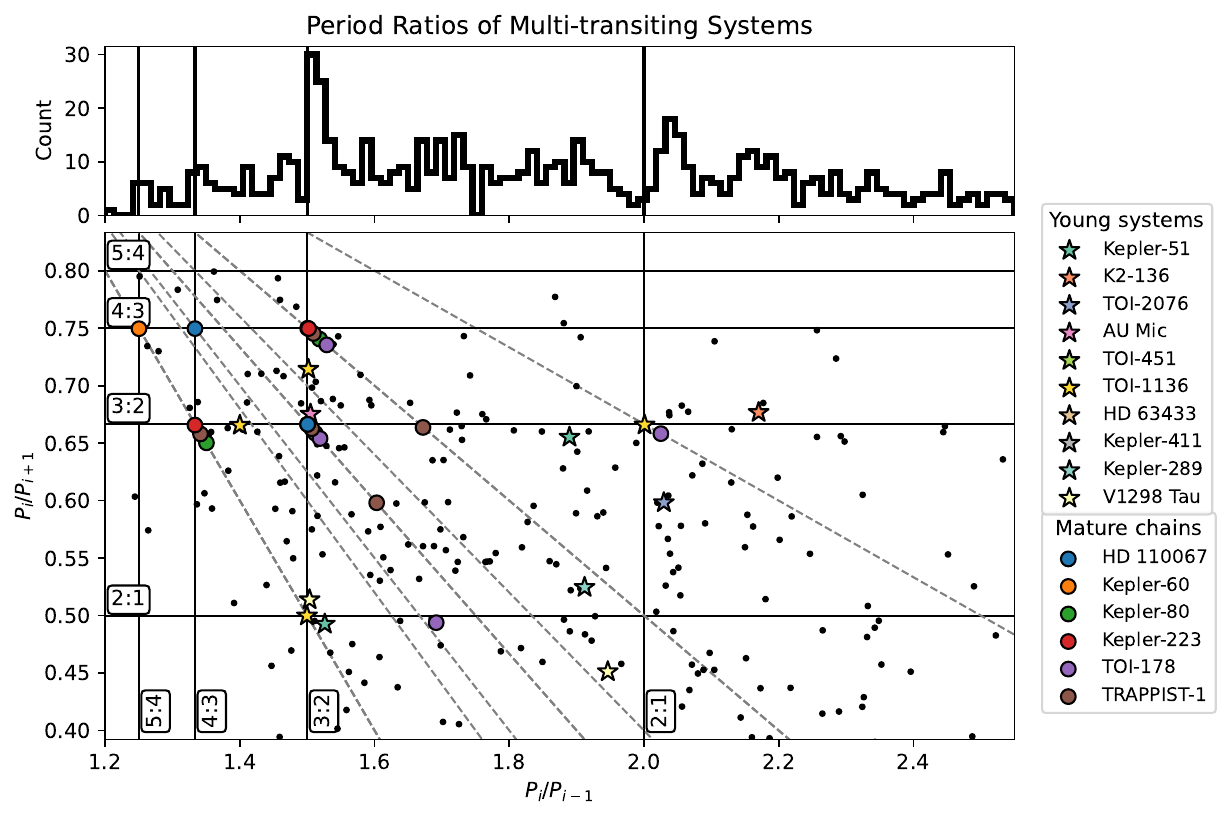}
    \caption{
    Period ratios of adjacent planets in transiting systems.
    The top panel shows the histogram of period ratios of all adjacent transiting planets in systems hosting two or more planets.
    The bottom panel plots the outer versus inner planet pair period ratios for adjacent trios of planets in systems with three or more planets. 
    The loci of first order two-body MMRs and zeroth-order three-body MMRs are indicated by solid black and gray dashed lines, respectively. 
    Young multi-planet systems analyzed by \citet{Dai2024} are indicated by stars. 
    Mature resonant chains, shown as large circles, lie along three-body MMRs. 
    Most multi-transiting systems eschew resonant configurations, but show a preference for period ratios just wide of the resonances.
    Data are taken from the NASA Exoplanet Archive \citep{exoplanet-archive-2013}.}
    \label{fig:pp-plane-observed}
\end{figure*}

\subsection{MMR deficit and the Pile-up}
\label{sec:intro:pileups}

The observed period spacings of super-Earths are intriguing. While there are a handful of instances where super-Earths are arranged in resonant chain configurations \citep[e.g.,][]{Godziewski2016,Mills2016,MacDonald2016,Leleu2021,Dai2023,Luque2023}, the vast majority of multi-planet systems show no preference for low-order mean-motion resonances (MMRs) between their constituent planets. Rather, the period ratios are distributed broadly. This runs against one's naive expectation if dissipative orbital migration has taken place in these systems \citep{Terquem2007,Cresswell2008,Izidoro2017}. 
There are, however,  peculiar excesses of pairs a couple percent wide of certain first-order MMRs
\citep[][see also Figure \ref{fig:pp-plane-observed}]{Lissauer2011,Fabrycky2014,SteffenHwang2015}.
If we define the proximity to resonance as
\begin{equation}
\label{eq:delta_def}
\Delta \equiv \frac{j-1}{j}\frac{P_{\rm out}}{P_{\rm in}} - 1\, ,
\end{equation}
where $j$ is an integer designating the particular first-order MMR and $P_{\rm in}$ and $P_{\rm out}$ are the inner and outer planets' orbital periods, respectively, one observes in Figure \ref{fig:pp-plane-observed} prominent peaks near the 3:2 and the 2:1 MMRs at $\Delta \sim 1-2\%$. Moreover, this excess appears to be much more prevalent among younger systems \citep{Dai2024}. Among systems younger than 100 Myrs, 7 out of 10 pairs are within a few percent of an exact MMR, compared to $\sim 15\%$ in mature systems.

Much has been written to explain these oddities.
To explain the paucity of MMRs, it is possible that super-Earths are formed {\it in-situ} and avoid  dissipative migration \cite[e.g.][]{Hansen2012}, or that stochastic torques from disk turbulence  \citep{Adams2008,Rein2012,GoldbergBatygin2023}  help migrating planets evade resonant capture, or that captured planets escape  resonance via eccentricity damping and over-stable libration
\citep{Goldreich2014}. However, the first path-way ignores the important role of gas as these planets are formed and the latter two pathways have been argued to be ineffective \citep{Batygin2017,Deck2015}.

Lately, popular attention has  shifted to the so-called ``breaking-the-chains" model \citep[e.g.][]{Izidoro2017,Izidoro2021,Li_OConnor_2025,Li2025,GoldbergPetit2025}. Here, super-Earths are captured into chains of MMRs while they are forming and migrating in gaseous disks, and later on, an event leads to dynamical instability in these closely-packed systems. The subsequent orbital excitations and mergers give rise to the largely feature-free period ratio distribution we see today. This model is able to broadly reproduce several aspects of the observed exoplanetary population \citep[see also][]{Goldberg2022,Li2025}. Moreover, if the instabilities occur on $100$\,Myrs timescale, it may also explain the  higher incidence of (near) resonant configurations found among younger systems \citep{Dai2024}.

Similarly, there have also been many proposals to explain the pile-ups wide of MMRs. Prolonged eccentricity damping pushes planet pairs apart \citep{LithwickWu2012,BatyginMorbidelli2013,Delisle2014}.  Due to resonant dynamics, for an initially flat period ratio distribution, this can create an excess of period ratios just wide of first-order MMRs and a corresponding deficit of ratios just narrow of the same MMRs. Hence this effected has been dubbed ``resonant repulsion" \citep{LithwickWu2012}. However,  it has been pointed out \citep[e.g.,][]{ChoksiChiang2023,Wu2024} that such an extensive eccentricity damping is inconsistent with the values of free eccentricity measured by transit timing  variations \citep{Lithwick2012,hadden_2014,Hadden2017,GoldbergBatygin2023}.

This brings us to another class of proposals for the near-resonant period ratio pile-ups. \citet{Chatterjee2015} first showed, numerically, that planetesimal scattering  can also repel a pair of planets \citep[see also][]{Ghosh2023}. The dynamics behind this behavior was elucidated by \citet{Wu2024}, where they argued that the plantesimals tend to absorb fractionally more energy than angular momentum, as they scatter off the two planets repeatedly. This works as dynamical friction and repels planet pairs, similar to direct eccentricity damping. However, it differs from pure damping in one key aspect: if planetesimals (which they dub ``ping-pongs") are not too low in mass, they may excite the planets' free eccentricities to the so-called equipartition values. They estimated that, to reproduce the observed pile-ups and the free eccentricities, the total mass of the ping-pongs are of order a few percent of the planet masses, and individual ping-pongs should be of order Mercury in mass.

\subsection{Our Proposal: ``rattle-and-break"}
\label{sec:intro:scattering}

In this work, we explore a way to explain both of the aforementioned observational oddities in one coherent framework. Our proposal combines ``ping-pong" repulsion with the ``breaking-the-chains" idea, and we name it the ``rattle-and-break" model.  Here, we discuss our motivations.

First, the ``breaking-the-chains" model faces a serious issue. The dissipative forces responsible for establishing resonant chains should leave the chains in exceptionally stable orbital arrangements. These are dynamical attractors that lie close to the elliptic equilibrium of the system's conservative dynamics \citep[e.g.,][]{Delisle2017}.\footnote{Technically, the dynamical configurations towards which resonant chains evolve are elliptic equilibria of an appropriately reduced phase space. These elliptic equilibria in the reduced phase space correspond to periodic orbits in a co-precessing reference frame \citep[see, e.g,][]{Gozdziewski2020}.} Such dynamical configurations should be exceptionally long-lived or even indefinitely stable \citep[e.g.,][]{poschel1989elliptic}. Numerical simulations by \citet{Li2025} confirm this picture: after trapping systems of six planets into resonant chains under the effects of dissipative migration, they report that the vast majority remain stable for the duration of their 100~{Myr} integrations.  This is in contrast with the findings of \citet{Izidoro2017,Izidoro2021}, who report unstable fractions of $50-90\%$. We discuss the origin of this discrepancy in Section \ref{sec:discussion:compare}. In short, in order for the ``breaking-the-chain" model to succeed, some additional dynamical mechanism is needed to trigger instabilities in primordial resonant chains.

 We suggest in this work that planetesimal scattering is one such dynamical mechanism. Scattering  low-mass bodies causes orbital repulsion and eccentricity excitation for the planets. Both dislodge the resonant chains away from their stable equilibrium. In closely-packed systems, even small displacements may lead, over time, to full-scale orbital instability.

 Previous works have also invoked planetesimal scattering to break resonant chains \citep{Li_OConnor_2025,GoldbergPetit2025,OgiharaKunitomo2025}, including primordial chain configurations among our Solar System's outer planets \citep{Morbidelli2007}.  We discuss the ways in which our work differs from earlier ones in \S \ref{sec:discussion:compare}.

In this work, we are interested in extending the concept of repulsion. While \citet{Wu2024} studied how a pair of planets are repelled by scattering, we extend it to a chain of resonant planets. Does scattering cause all pairs in a chain expand in spacing, or just some? Can such a picture explain the observed pile-up? In cases where scattering causes wholesale destruction, we are also interested in understanding how to interpret the apparent demise of resonant pairs over 100\,Myr timescales. Overall, we wish to establish specific requirements for the ``rattle-and-break" model, if it is indeed the dominant mechanism through which planetary systems reach their mature state.

The plan of the paper is as follows: 
in Section \ref{sec:simulations} we describe the sets of $N$-body simulations we use to explore the effects of planetesimal scattering on resonant chains. Section \ref{sec:results} summarizes the results of our simulations. We discuss  various issues in Section \ref{sec:discussion} and conclude in Section \ref{sec:summary}.

\section{Rattling the Chains: Set-up}
\label{sec:simulations}
 We run a series of $N$-body simulations of resonant chain systems that include populations of low-mass planetesimals. 
All simulations are conducted using the \texttt{trace} hybrid integrator
\citep{trace_lu_2024} implemented in the \texttt{rebound} package \citep{Rein2012rebound}.
Briefly, this integrator works by switching from a symplectic, Wisdom-Holman style integrator \citep{wh1991} to a high-accuracy Bulirsch-Stoer  integrator whenever bodies undergo close encounters below some user-specified threshold distance. In our simulations, we set the close encounter threshold for integrator switching to be 5 mutual Hill radii.
The integrator time step is set to $1/30$ times the initial orbital period of the innermost planet. In all simulations, small bodies' gravitational interactions with the planets are accounted for, but their mutual gravitational interactions are ignored by setting the \texttt{rebound} simulation the parameter \texttt{testparticle\_type} to 1. 

In all simulations, we take the mass of the star to be $1\,M_\odot$ and the initial semi-major axis of innermost planet to $0.1\,\mathrm{AU}$.  All objects are assigned a bulk density of 1 g/$\mathrm{cm}^3$.\footnote{This gives an $8M_\oplus$ planet a radius of $3.5 R_\oplus$, larger than that ($1.5 R_\oplus$) adopted in \citet{Wu2024}. Our larger radii lead to shorter accretion times and a less efficient repulsion per unit planetesimal mass.} Collisions are resolved by merging colliding particles while preserving their total momentum.

The initial conditions of our simulations are described below and summarized in Table \ref{tab:sim_summary}. We run two sets of simulations with different orbital architectures. Our first set of simulations are based on the HD 110067 system, a resonant chain of 6  super-Earths whose adjacent planet pairs, from innermost to outermost, have period ratios of 3:2, 3:2, 3:2, 4:3 and 4:3. The planets' masses are $(m_1,...,m_6) = (7.2, 7.9, 10.7, 4.9, 6.3, 10.5)M_\oplus$.  Our second set of simulations are based on the Kepler-223 system, a resonant chain of 4 super-Earths with period ratios, from the innermost to the outermost pair, of 4:3, 3:2 and 4:3. The planets' masses are $(m_1,m_2,m_3,m_4) = (6.6,4.5,7.1,4.3)M_\oplus$.  We assume these mature resonant chain systems' orbital configurations are similar to the initial configurations of super-Earth systems more broadly, as proposed in the ``breaking-the-chain" model.

Simulations are divided into two phases: an initial setup phase designed to mimic the outcome of migration and eccentricity damping in protoplanetary disks, and a main phase where the system evolves purely under the influence of mutual gravitational interactions and collisions.

The set-up phase starts with the planets initialized in a preliminary resonant equilibrium configurations\footnote{These preliminary equilibria are chosen, somewhat arbitrarily, such that $e_1 = 0.013$ for HD 110067 and $e_1 = 0.016$ for  Kepler-223.} computed analytically using equations of motion from the analytic model described by \citet{Delisle2017} and implemented in the \texttt{celmech} code \citep{Hadden2022}. 

 We also embed a number of small bodies within the planetary region.
All small bodies in a given simulation have the same mass, $m_s$, measured in unit of Pluto mass ($M_{\rm Pluto}$).  We also vary the total mass of small bodies, measured as a fraction $f$ of the planets' total mass. We choose $f= 3\%$ or $10\%$. These fractions amount to up to a few Earth masses of small bodies. The small bodies are initialized on circular orbits with semi-major axes chosen uniformly from the range $a\in [a_1 - 10R_\mathrm{H,1},a_N + 10R_\mathrm{H,N}]$ where $a_1$ and $R_\mathrm{H,1}$ (resp., $a_N$ and $R_\mathrm{H,N}$)  are the semi-major axis and Hill radius of the innermost (resp., outermost) planet. Small bodies' initial semi-major axes are excluded from regions spanning $a_i \pm 3R_\mathrm{H,i}$ around each planet's orbit. Small body orbital inclinations are drawn from a Rayleigh distribution with scale parameter $\sigma = 0.01$ and their angular orbital elements are drawn at random from uniform distributions.

The addition of small bodies perturb the resonant chains. We opt to relax the chains to  a new dynamical equilibrium by applying  migration and eccentricity damping forces. Specifically, we use the \texttt{modify\_orbits\_direct} operator implemented in  the \texttt{reboundx} package \citep{Tamayo2020reboundx} to apply semi-major axis and eccentricity damping to all planets. The rates of semi-major axis damping are chosen so that  $d\ln(a_1  /a_6)/dt = 10^6P_{1}$, $d\ln(a_1  /a_i)/dt = 0$ for $i = 2,...,5$, and $\frac{d}{dt}\sum_{i}\frac{m_i}{a_i} = 0$. The latter condition ensures the system's total energy is approximately conserved and is imposed for numerical convenience to avoid  large-scale migration of the system. This has no impact on the resulting resonant dynamics, which depend only on planets' relative migration rates. Planets' eccentricities are all damped at a constant rate so that $d\ln e_i/dt = -10^6P_{1} / K$ where $K=10$ or $100$. 
The small bodies' eccentricities are also damped, at a rate of $d\ln e_s/dt = -10^3P_{1}$. Damping forces are applied for a duration of $10^5P_1$. 

During this relaxation period, the planets re-settle into stable resonant configuration. Afterwards, damping forces are  turned off and the main phase of our simulations commence.  Each simulation is integrated up to a maximum time of $10^{10}P_1\approx320~\mathrm{Myr}$. During the simulations, any bodies that become gravitationally unbound are periodically removed and the system is re-centered at the center of mass of the remaining bound particles.

\begin{table*}[ht]
\centering
\caption{Results of rattling the chains. 
No instability times ({marked by} `--') are recorded for simulations that remained stable over
the full $320\,{\rm Myr}$ duration of our simulations.
} 
\label{tab:sim_summary}
\begin{tabular}{lccccc}
\hline
\hline
Simulation Name & Resonances & $m_s~[M_\mathrm{Pluto}]$ & $K$ & $f$ & Instability time [Myr] \\
\hline
\texttt{hd110067\_ms3\_K10\_f0.1  } & 3:2, 3:2, 3:2, 4:3, 4:3 & 3 &  10 & 0.1  & 0.2 \\
\texttt{hd110067\_ms3\_K100\_f0.1 } & 3:2, 3:2, 3:2, 4:3, 4:3 & 3 & 100 & 0.1  & 0.2 \\
\texttt{hd110067\_ms3\_K10\_f0.03 } & 3:2, 3:2, 3:2, 4:3, 4:3 & 3 &  10 & 0.03 & 7.4 \\
\texttt{hd110067\_ms3\_K100\_f0.03} & 3:2, 3:2, 3:2, 4:3, 4:3 & 3 & 100 & 0.03 & 6.1 \\
\texttt{hd110067\_ms2\_K10\_f0.1  } & 3:2, 3:2, 3:2, 4:3, 4:3 & 2 &  10 & 0.1  & 0.2 \\
\texttt{hd110067\_ms2\_K100\_f0.1 } & 3:2, 3:2, 3:2, 4:3, 4:3 & 2 & 100 & 0.1  & 4.1 \\
\texttt{hd110067\_ms2\_K10\_f0.03 } & 3:2, 3:2, 3:2, 4:3, 4:3 & 2 &  10 & 0.03 &  9.2 \\
\texttt{hd110067\_ms2\_K100\_f0.03} & 3:2, 3:2, 3:2, 4:3, 4:3 & 2 & 100 & 0.03 &  8.9 \\
\texttt{hd110067\_ms1\_K10\_f0.1  } & 3:2, 3:2, 3:2, 4:3, 4:3 & 1 &  10 & 0.1  & 13.4 \\
\texttt{hd110067\_ms1\_K100\_f0.1 } & 3:2, 3:2, 3:2, 4:3, 4:3 & 1 & 100 & 0.1  & 46.3 \\
\texttt{hd110067\_ms1\_K10\_f0.03 } & 3:2, 3:2, 3:2, 4:3, 4:3 & 1 &  10 & 0.03 & 100.8 \\
\texttt{hd110067\_ms1\_K100\_f0.03} & 3:2, 3:2, 3:2, 4:3, 4:3 & 1 & 100 & 0.03 & -- \\
\hline
\texttt{k223\_ms3\_K10\_f0.1   }    & 4:3, 3:2, 4:3            & 3 &  10 & 0.1  & -- \\
\texttt{k223\_ms3\_K100\_f0.1  }    & 4:3, 3:2, 4:3            & 3 & 100 & 0.1  & -- \\
\texttt{k223\_ms3\_K10\_f0.03  }    & 4:3, 3:2, 4:3            & 3 &  10 & 0.03 & --  \\
\texttt{k223\_ms3\_K100\_f0.03 }    & 4:3, 3:2, 4:3            & 3 & 100 & 0.03 & --  \\
\texttt{k223\_ms1\_K10\_f0.1   }    & 4:3, 3:2, 4:3            & 1 &  10 & 0.1  & -- \\
\texttt{k223\_ms1\_K100\_f0.1  }    & 4:3, 3:2, 4:3            & 1 & 100 & 0.1  & --  \\
\texttt{k223\_ms1\_K10\_f0.03  }    & 4:3, 3:2, 4:3            & 1 &  10 & 0.03 & --  \\
\texttt{k223\_ms1\_K100\_f0.03 }    & 4:3, 3:2, 4:3            & 1 & 100 & 0.03 & --  \\
\hline
\end{tabular}
\end{table*}

\section{ Rattling the Chains -- 
Consequences}
\label{sec:results}

We present the results of our numerical simulations for the two analogue groups.  Although they experience similar amounts of planetesimal scattering, the final outcomes are different. 

\subsection{the HD 110067 analogues}
\label{sec:results:hd110067}
\begin{figure}
    \centering
    \includegraphics[width=\linewidth]{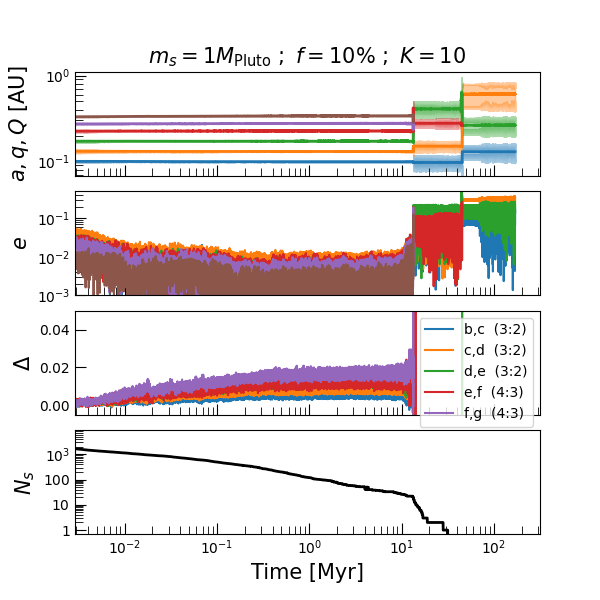}
    \caption{
    How a resonant chain can be rattled free by planetesimal scattering. Summary of simulation \texttt{hd110067\_ms1\_K10\_f0.1}. The top panel shows the semi-major axes (solid lines) and apocenter-pericenter range (shaded regions) of each planet versus time. The next panel shows planets' orbital eccentricities. The third panel shows adjacent planet pairs' distance from resonance, Equation \eqref{eq:delta_def}. The fourth panel shows the number of surviving bound planetesimals. The system undergoes a orbital instability after $13\,{\rm Myr}$, resulting in planet mergers, followed but another instability and merger around $40\,{\rm Myr}$, leaving a three-planet system. 
    }
    \label{fig:hd110067_evolution_example}
\end{figure}

 Figure \ref{fig:hd110067_evolution_example}  shows the evolution of run \texttt{hd110067\_ms1\_K10\_f0.1}. As planetesimal scatterings
removes orbital energy from the planet ladder and transfers angular momentum from the inner to the outer parts, the planets'  eccentricities are reduced and all planet pairs  spread apart.  
The values of $\Delta$ increase from nearly zero to values of $\sim 0.01$, as shown in  
Figure \ref{fig:hd110067_evolution_example}. Orbital instabilities set in after the system is no longer protected by the resonances, with  orbit crossings and subsequent mergers ultimately reducing the number of planets from six to three.  Most of the scattering planetesimals have been absorbed by the planets after a mere $10^5$ yrs.

\begin{figure*}
    \centering
    \includegraphics[width=0.9\linewidth]{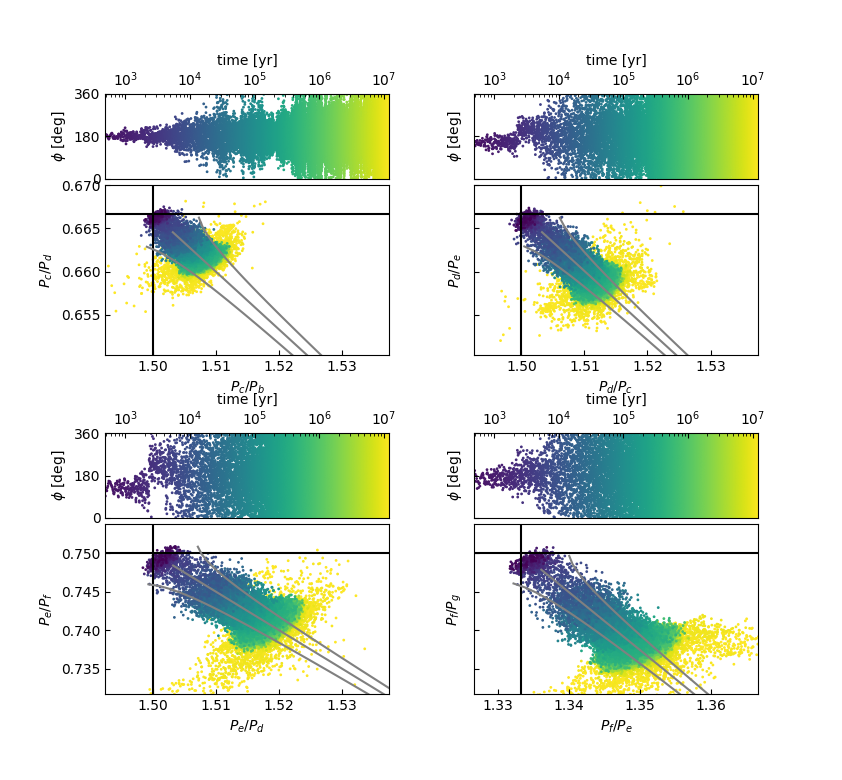}
    \caption{Same as Fig. \ref{fig:hd110067_evolution_example} but now emphasize evolution in the three-body resonances, with each group depicting one adjacent trio in the system. In each panel, the top sub-panel shows the time evolution of a trio's three-body resonance angle, $\phi$ (Equation \ref{eq:phi_def}). Bottom sub-panels show the outer planet pair's period ratio versus the inner pair's, colored according to simulation time. The centers and libration widths of each three-body resonance, computed  according to Equation C21 of \citet{Lammers2024},  are indicated by gray lines. Planetesimal scattering causes adjacent planet pairs spread wide of nominal two-body resonance while closely following three-body resonances in the period-ratio plane until an orbital instability occurs after $13\,\rm{Myr}$.
    }
    \label{fig:hd110067_3br_example}
\end{figure*}

 Repulsion of planet pairs is generic. We show in Figure \ref{fig:hd110067_3br_example} the evolution of the same simulation  but in planes of orbital period ratios.  Every adjacent pair spread wide of resonance.  We also notice that adjacent planet trios closely follow the loci of three-body resonances in the period ratio plane.  Such  behavior is characteristic of resonant chain systems  when subject to eccentricity damping \citep[e.g.,][]{Goldberg2021}. Indeed, smooth eccentricity damping alone should move the entire system along a one-parameter family of equilibrium configurations \citep{Delisle2017}, one that maintains zero libration in all three-body angles,
\begin{equation}
\label{eq:phi_def}
    \phi_{i} = \frac{p_{i+1}\lambda_{i+2}-(p_{i+1}+p_{i}-1)\lambda_{i+1} + (p_i-1)\lambda_i}{{\rm gcd}(\{p_{i+1},p_{i+1}+p_{i}-1,p_i-1\})},
\end{equation}
where $\lambda_i$ is the mean longitude of the $i$th planet, and where $p_1=p_2=p_3 =3$ and $p_4=p_5 = 4$ for our HD 110067 analogues.

 In our simulation, however, these angles do not remain in libration for long. After $\sim 10~\mathrm{kyr}$, the angles for all but the innermost trio of planets start circulating. We attribute the resonance breaking to the inherent ``graininess" of planetesimal scattering \citep{Nesvorny2016}. While these scatterings produce, on average, an eccentricity damping effect through dynamical friction and collisions,  individual encounters and collisions superimpose random perturbations. As the three-body resonances become narrower and weaker further away from two-body commensurability (see Figure \ref{fig:hd110067_3br_example}), these random perturbations can more easily disrupt the three-body resonances.

Despite the abrupt disruption of three-body angle libration shown in Figure \ref{fig:hd110067_3br_example}, the planets' subsequent orbital period evolution still closely follows the three-body resonances, i.e., the regions delimited by gray lines in the figure. Thus, the three-body resonances remain dynamically important, exhibiting an apparent ``stickiness" that guides the evolution of period ratios as planetesimal scattering continues to push planet pairs wide of MMR. Examining the evolution of resonant angles at a higher time resolution than shown in Figure \ref{fig:hd110067_3br_example} reveals that they often intermittently switch between libration and circulation. By approximately $\sim 1\,\mathrm{Myr}$, the planets' orbital migration has essentially ceased because almost all planetesimals have been accreted into planets and their orbital evolution is dictated mainly by their mutual interactions, which eventually leads to an orbital instability.

\begin{figure*}
    \centering
\includegraphics[width=0.9\linewidth]{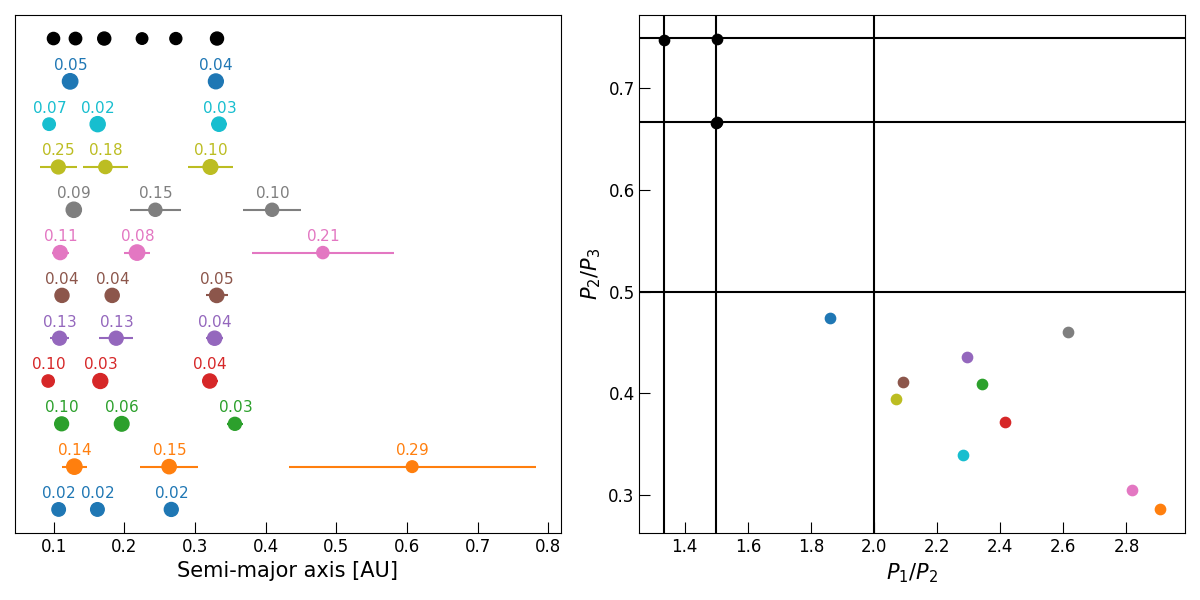}
    \caption{
    The initial (in black) and final orbital architectures of HD 110067 analogue systems that underwent dynamical instabilities. The left-hand panel shows the orbital semi-major axes of planets, along with their apocenter and pericenter distances, indicated by error bars. Average eccentricities, computed over the last 10\% of simulation time, are indicated above each planet. The right-hand panel shows, for systems with more than two surviving planets,  where they fall in the period ratio-period ratio plane. Once the instabilities occur, all MMRs are disrupted.
    }
\label{fig:unstable_architecture}
\end{figure*}

Our other simulations of HD 110067 show dynamical behavior similar to that seen in Figures \ref{fig:hd110067_evolution_example} and \ref{fig:hd110067_3br_example}.  By the end of our integrations, all but one of the HD 110067 analogue simulations  have experienced a dynamical instability, here taken to mean a  fractional change in any planet's semi-major axis of $>10\%$. The timescales on which these instabilities occur range from $0.2\,\mathrm{Myr}$ to $100\,\mathrm{Myr}$. Simulations with more massive small bodies (larger $m_s$)  and greater total masses of small bodies ($f$) tend to experience earlier instabilities (see Table \ref{tab:sim_summary}).   In contrast, the value of $K$ is  unimportant. The final orbital configurations are shown in Figure \ref{fig:unstable_architecture}. Unstable systems are eventually reduced via planet-planet mergers to systems of three or, in one instance, two planets. These systems tend to have high eccentricities of order $\sim 0.1$ (along with inclinations of order $\sim 0.05$), and wide period spacings, typically beyond the 2:1 MMR.

\subsection{the Kepler-223 analogues}
\label{sec:results:kep223}
\begin{figure}
    \centering
    \includegraphics[width=0.95\linewidth]{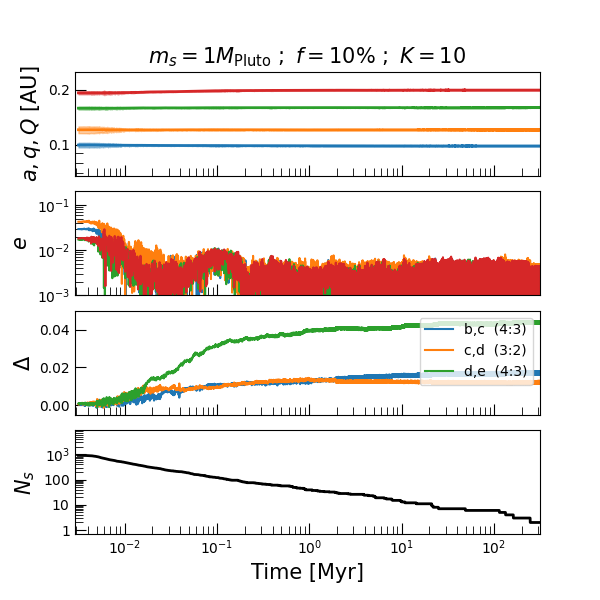}
    \caption{
    Same as Figure \ref{fig:hd110067_evolution_example}, but for an Kepler-223 analogue (\texttt{k223\_ms1\_K10\_f0.1}).
    In contrast with HD 110067 simulations, all simulations of Kepler-223 remain dynamically stable for at least $320\,{\rm Myr}$. 
    }
    \label{fig:k223_example}
\end{figure}

\begin{figure*}
    \centering
    \includegraphics[width=\linewidth]{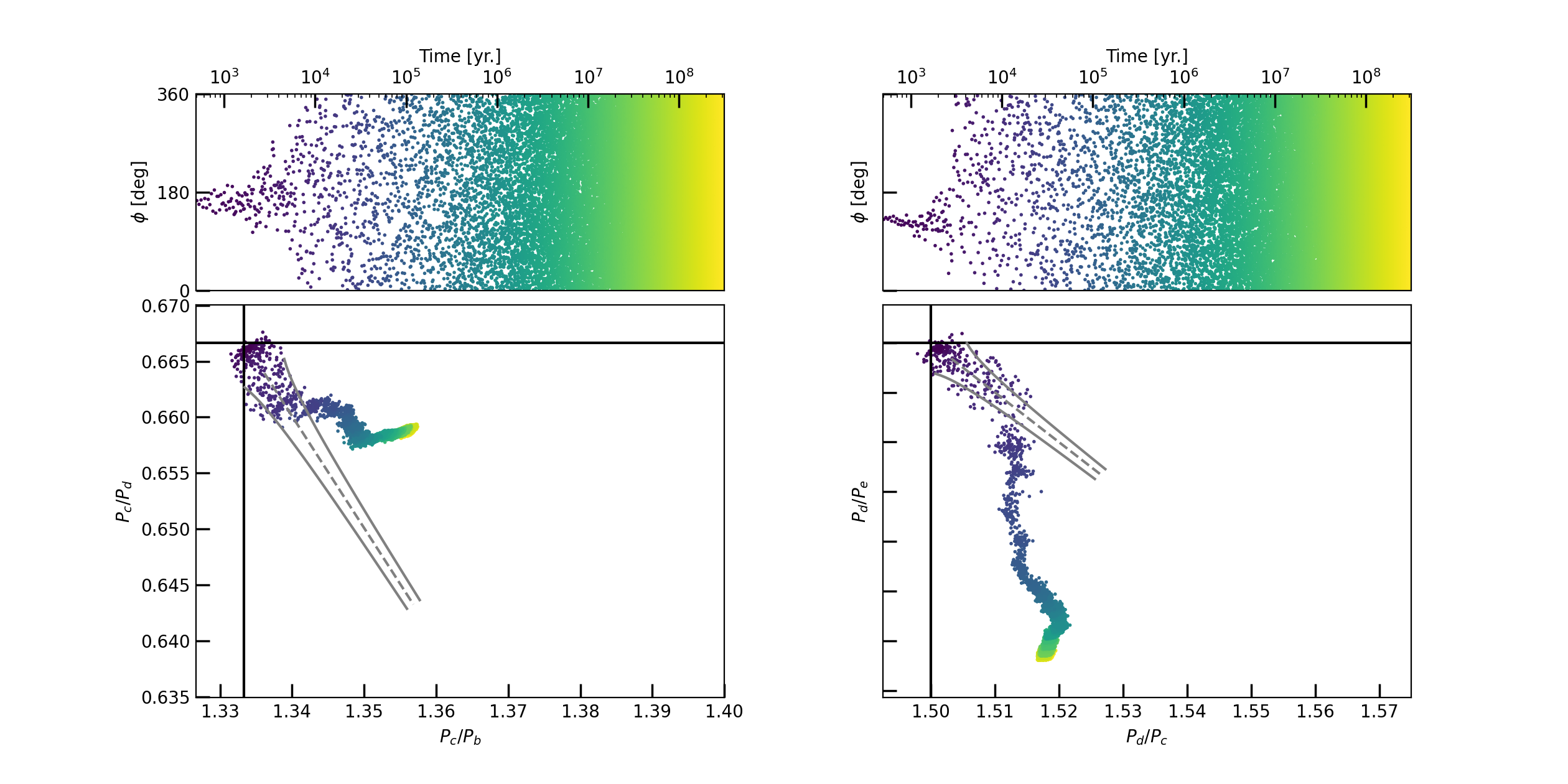}
    \caption{
    Similar to Figure \ref{fig:hd110067_3br_example}, but for the 
    \texttt{k223\_ms1\_K10\_f0.1} simulation  (also see Figure \ref{fig:k223_example}). Initially, planetesimal scattering spreads planet pairs wide of resonance while following three-body resonances, until  escape at $\sim 10^4~{\rm yr}$. Subsequently, continued scattering rapidly drives the outermost planet pair, $(d,e)$, apart  along with some additional divergent migration of the innermost pair, $(b,c)$ while the middle planet pair, $(c,d)$ remains at roughly fixed orbital separation. Three-body resonance centers and maximal libration widths, computed  according to the procedure described in \citet{Lammers2024}, are indicated by gray dashed and solid lines, respectively. 
    }
    \label{fig:k223_3br_example}
\end{figure*}

In contrast with simulations of HD 110067, all  analogues of the Kepler-223 system remain dynamically stable, even after planetesimal scattering disrupt the resonant chains. Figures \ref{fig:k223_example} and \ref{fig:k223_3br_example} summarize the evolution of the \texttt{k223\_ms1\_K10\_f0.1} simulation. As in the case of the \texttt{hd110067\_ms1\_K10\_f0.1} simulation presented in Section \ref{sec:results:hd110067}, planetesimal scattering  increases the period ratios wide of the original MMRs, initially  maintaining the three-body resonances.  However, as seen in Figure \ref{fig:k223_example}, the outermost pair of planets  (planets d,e)  begins to diverge more rapidly than the other pairs after $\sim10~\mathrm{kyr}$. This  pulls adjacent planet trios  away from three-body resonances. The system eventually exhausts its supply of planetesimals and further migration essentially halts after $\sim 1\,{\rm Myr}$. Unlike the HD 110067 analogues, we find that the systems remain dynamically stable until the end of the integration  (i.e., 320 Myrs). We suggest this difference may arise from the smaller number of planets in Kepler-223 (see \S \ref{sec:discussion}).

 It is interesting to note that \citet{Moore2013} also used Kepler-223 as a case study for planetesimal scattering. Their principal aim was to constrain the  maximum amount of mass the system could have scattered while  retaining its observed resonant configuration.
Their results are qualitatively similar to ours, including the spreading of planet pairs wide of MMRs and the absence of dynamical instabilities, though their integrations stop at $\sim 10^4$\,yrs.

\begin{figure}
    \centering
    \includegraphics[width=0.95\linewidth]{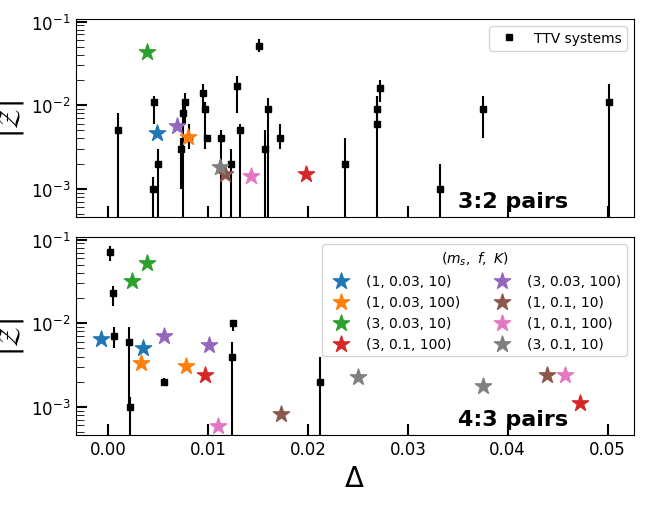}
    \caption{
    Magnitudes of combined complex eccentricities (Equation \ref{eq:zcomb}) versus distance to MMR, $\Delta$  (Equation \ref{eq:delta_def}), for  the Kepler-223 analogues and for 
    observed TTV planet pairs near the 3:2 (top panel) and 4:3 (bottom panel) resonances. Data are taken from \citet{Hadden2017}.
    The same correction from osculating to free complex eccentricities 
    used by \citet{Hadden2017} is applied to our simulated data. 
    }
    \label{fig:ZvsDelta}    
\end{figure}

Figure \ref{fig:ZvsDelta} summarizes the final orbital arrangements of our Kepler-223 analogues. In four out of eight simulations, the three-body resonances of both  the inner and the outer planet trios  are entirely disrupted by planetesimal scattering, similar to the simulation 
shown in Figure \ref{fig:k223_3br_example}. The remaining four simulations, mostly those with a smaller fraction of planetesimal masses ($f=0.03$), exhibit various combinations of two- or three-body MMRs among at least some of the constituent planets. Fig. \ref{fig:ZvsDelta} shows the final distances to 2-body resonances ($\Delta$) for all planet pairs. 
 We also present the pairs' free eccentricities, as measured in complex notation and with the following combination,
\begin{equation}
    \mathcal{Z} =
    \frac{
    f_j e_{\rm in}\exp\left[\mathrm{i}\varpi_{\rm in}\right]
    +
    g_j e_{\rm out}\exp\left[\mathrm{i}\varpi_{\rm out}\right]}
    {\sqrt{f_j^2+g_j^2}},
    \label{eq:zcomb}
\end{equation}
where $f_j$ and $g_j$ are coefficients that depend on the particular $j$:$j-1$ resonance  and $\varpi$ the pericenter angles \citep[see, e.g.][]{Hadden2016}. This form is adopted because it is directly measurable from transit-time-variations (TTVs). In Fig. \ref{fig:ZvsDelta}, we  plot values of $\Delta$ and ${\cal Z}$ for the observed TTV pairs (not only in Kepler-223), as obtained by \citet{Hadden2017}.
We can broadly reproduce the observed free-eccentricities, though our values tend to be on the lower end of the measurements. It may indicate that the ping-pongs are of order Mercury mass \citep{Wu2024}, some $\sim20$ times more massive than  is assumed here.

\section{Discussion}
\label{sec:discussion}

\subsection{Comparison to observed trends}
\label{sec:discussion:observations}

{The central tenet of the ``breaking-the-chains" hypothesis \citep{Izidoro2017,Izidoro2021} is that the current orbital configurations of close-in planets arise from dynamical instabilities among initially compact resonant chains. Here we test one particular version of this hypothesis in which scattering of low-mass planetesimals (`ping-pongs') rattles planet pairs away from the protection of MMRs. We find that this leads to widening of the planetary spacings and, in some cases, full-scale dynamical instabilities. While our limited numerical explorations prevent us from making a full statistical comparison with the observed population, we can nevertheless draw the following conclusions.}

If every super-Earth system has experienced similar planetesimal scattering to that simulated here, it can explain a number of the  observed trends in these systems' orbital architectures:
 \begin{enumerate}
\item  Scattering  small bodies always repels planet pairs away from MMRs. This is true not just for pairs of planets but for chains of resonant planets as well.  This explains the excess of planet pairs at $\Delta > 0$.

\item We find that scattering is largely finished after a mere $10^4-10^5$ yrs. Accretion of the ping-pongs into the planets truncate the scattering process \citep[see Fig. 7 of][]{Wu2024}. As a result of this early rattling, we expect most of the young ($<100$Myrs) systems to reside near, but not in, resonances, as found by \citet{Dai2024}. Similarly, these young systems may not
lie along lines of strong three-body resonances,  as is indeed the case with the exception of TOI-1136 (Fig. \ref{fig:pp-plane-observed}). %

\item For rattled chains that do not undergo dynamical instabilities (e.g., Kepler-223 analogues),  neighboring planets should sit just wide of first-order MMRs, by an amount in $\Delta$ that is determined by the total fractional mass in the ping-pongs, and with a free eccentricity that scales as the squar root of the individual ping-pong mass \citep[also see][]{Wu2024}.  The observed values are both of order a few percent,  and suggest that the planetesimals are of order a few percent in total fractional mass, and are similar to Mercury in individual mass.

\item  Almost all our HD 110067 analogues are destabilized after  scattering, on timescales that range from $10^5$ - $10^8$ yrs. The  timescale appears to depend on individual ping-pongs' masses. If it is indeed true that the incidence of near-resonant systems decay over $10^8$-$10^9$ yrs  \citep{Dai2024}, this would point to ping-pong masses of order the Pluto mass. This is about an order of magnitude lower than the above constraint from free-eccentricities. 
In addition, such systems should harbor fewer planets that are more widely spaced, with a broad distribution in period-ratios and with eccentricities (and inclinations) of order $10\%$ (Fig. \ref{fig:unstable_architecture}). This may explain the high eccentricity population observed in the TTV sample \citep{Lithwick2012,hadden_2014}, 
     as well as the high eccentricity planets inferred from transit durations \citep{VanEylen2015,Xie2016PNAS,VanEylen2019}.
 \end{enumerate}
 
There is tension between Mercury-mass planetesimals suggested by TTV eccentricity measurements and the Pluto-mass planetesimals required to produce the $\sim 100\,\rm{Myr}$ dynamical lifetimes.  It may be somewhat alleviated if  most primordial chains are less dynamically delicate than HD 110067.  

In summary, rattling  resonant chains with small bodies can leave some pairs wide of MMRs while also breaking  chains on the requisite timescales. We do have a discrepancy in estimating the ping-pong mass: while the magnitude of free eccentricities require Mercury-sized bodies, the instability timescale prefers Pluto-sized ones.

\subsection{A Dichotomy of Systems?}
\label{subsubsec:dichotomy}

In our simulations, we observe that the 6-planet systems (HD 110067 analogues) commonly undergo drastic dynamical reorganizations, while the 4-planet systems (Kepler-223 analogues) remain dynamically stable, at least until the end of our integrations (320 Myrs). These two types of systems are initialized with similar planet spacing and our experiments include a range of values in $K$, $f$ and $m_s$  (Table \ref{tab:sim_summary}).  This indicates that the number of planets in these closely packed systems is important for stability. This is reasonable:
\citet{Quillen2011,Petit2020,Rath2022,Lammers2024} have all argued that overlap among three-body resonances among a trio of planets is the underlying cause for instability in multiple systems, even those with more than three planets. Having a larger number of planets simply provides more trio combinations and therefore more chances for instability. 

Although our simulations are not exhaustive, we clearly observe the following  behavior: all near-resonant pairs are dismantled when dynamical instability occurs. In closely packed systems, local orbital instabilities quickly inflame into global instabilities, and almost all planets  experience orbit crossings and mergers. The final period-ratio distribution is a continuum. Consequently, near-MMR pairs are only retained in systems that avoid instability. These systems tend to be mildly eccentric and inclined, in contrast to the much more dynamically heated ones. They explain the excess of pairs wide of MMRs and may still bear memories of the original rattlers. 
Such a dichotomy may also be present in the works of other groups, but has only been explicitly reported by \citet{GoldbergPetit2025}. They analyzed stable and unstable systems in their simulations separately and call this dichotomy `a fundamental prediction of the breaking-the-chains model'.  There may exist some  observational support for this dichotomy. First, there is the original ``Kepler dichotomy",  an excess of single transiting planets relative to expectations given the multiple-transiting population \citep{Lissauer2011,Ballard2016}. This dichotomy suggests the presence of a closely-packed and well aligned population along with a population exhibiting wide orbital separations and high mutual inclinations. The first population may be primordial, while the second one may be the aftermath of dynamical instabilities \citep{Johansen2012}. Second, transit duration \citep{VanEylen2015,Xie2016PNAS,VanEylen2019} and TTV \citep{Lithwick2012,hadden_2014} studies reported the co-existence of both  low-eccentricity and a high-eccentricity populations. Third, \citet{Leleu2024}
 suggested that sub-Neptunes are on average more massive when they are further away from the MMRs ($\Delta > 0.05$). This can be explained if these planets are the results of planet mergers \citep{Li2025}.  

 Another natural prediction of the ``breaking-the-chains" model is that resonant or near-resonant pairs (at least those that are likely to be primordial) only occur with other such pairs. Fig. \ref{fig:pp-plane-observed} shows that this is not always true. However, \citet{Jiang2020,Dai2024} both pointed out that near-resonant pairs do tend to congregate. In particular, \citet{Dai2024} found that the chance of finding such a pair (defined to be $-0.015 \leq \Delta \leq + 0.03$) rises from $\sim 15\%$ in average system to $\sim 30\%$ in systems with already one known pair. This may be considered a partial success but more works are needed.

 Another prediction is that the young systems should closely resemble the primordial chains. Interestingly, with the exception of TOI-1136, none of the near-resonant chains identified by \cite{Dai2024} 
have more than four planets or possess any pair with a spacing closer than 3:2. 
In contrast, to successfully reproduce the observed period ratio distribution, one requires many primordial chains to be more closely spaced than even our HD 110067 analogues. For instance, more than half of  the planet pairs in the initial chains simulated by \citet{Izidoro2021} are closer than 4:3.  This should be further investigated.

\subsection{Comparison to previous works}
\label{sec:discussion:compare}

A number of studies have invoked scattering as a mechanism to break resonant chains. We compare our work against these, focusing on the different physical ingredients.

The first ingredient we consider is the mass of the scattering bodies. Many works have started simulations from planetary embryos that subsequently conglomerate and migrate in gas disks \citep[e.g.][]{Izidoro2017,Izidoro2021,OgiharaKunitomo2025,GoldbergPetit2025}.  By computational necessity, these embryos are initialized with fairly large masses, $\sim 0.1 M_\oplus$ or above.\footnote{\citet{Izidoro2021} started with lower mass seeds at $0.01 M_\oplus$. But due to high pebble fluxes in their simulations, even  `failed' embryos grow to masses of order $M_\oplus$.} As these embryos grow into super-Earths, they are captured into resonant chains by dissipative migration. These chains are later disrupted -- \citet{Izidoro2021} reported instability rates as high as $90\%$. While \citet{Izidoro2021} did not identify the mechanisms responsible for instabilities, we suspect, based on multiple figures in their work, that they result from the numerous left-over embryos that fail to grow into super-Earths.  These embryos, of order $\sim0.1-1 \,M_\oplus$ in mass, can easily disrupt the resonant chains. This supposition is supported by \citet{OgiharaKunitomo2025},  who found that system stability is best correlated with the total mass in such `failed' embryos.  Similarly, \citet{GoldbergPetit2025} showed that disturbances from low-mass embryos dislodge resonant chains (see, e.g., their Figure 15).

In contrast to the $\sim0.1-1\,M_\oplus$ Earth-mass perturbers in the above works, here we suggest that individual perturbers are much lower in mass, by about a factor of $\sim 100$. Such low-mass bodies can gently push planet pairs  wide of MMRs and impart modest  eccentricities and inclinations, explaining the observed near-resonant population.\footnote{\citet{OgiharaKunitomo2025} also recorded an excess outside $2:1$ MMR in one of their simulations (their Fig. 3). However, they give no explanation for this feature.} In contrast, Earth-mass rattlers do not provide sufficient dynamical friction to explain the near-MMR pile-ups. Instead, encounters with such massive perturbers  typically trigger large-scale orbital instabilities. As a result, theories with massive perturbers  predict that the current systems are either deeply in resonance (due to a total lack of encounters), or have lost all memories of the initial condition \citep[see, e.g.][]{GoldbergPetit2025}.

We now turn to the question of scattering bodies' source location. In this work, we have assumed a local population of planetesimal of currently-unknown origin. In the aforementioned works, left-over embryos that failed to grow and migrate are generally found just outside  primoridal chains' orbital regions.  \citet{Li_OConnor_2025} explored a different scenario, in which scattering planetesimals originate from a reservoir much further out in the system. The planetesimals fly  through the inner region on nearly parabolic orbits, presumably under the influence of outer massive perturbers.\footnote{ Reasonable doubts exist as to whether these mechanisms can consistently send inwards the massive population as required by \citet{Li_OConnor_2025}. Ejections could well occur earlier and interrupt the supply.}

While any scatterers may induce dynamical instability, only local (or nearly local) scatters, we suggest, can repel planet pairs apart. This is because only a local, quasi-circular planetesimal population can provide sufficient eccentricity damping through dynamical friction. Distant planetesimals on nearly parabolic orbits are too dynamically hot to provide the  eccentricity damping required to systematically drive resonant pairs apart. Our limited numerical experiments (not shown here) support this suggestion. 

In summary, we advocate for  a version of the ``breaking-the-chains" scenario in which low-mass planetesimals repel initially-resonant planet pairs wide of MMRs. In some systems this leads to wholesale instability. Repulsion requires planetesimals that are low in mass and are largely local in origin. These inferences are not valid, however, if repulsion and instability arise from two different processes.

 In our scenario, the pile-ups wide of MMRs result from the fact that all pairs were initially in MMRs. This differs from the original resonant repulsion story \citep[e.g.,][]{Lithwick2012,Wu2024}, where the initial distribution of the period ratios is flat, and the excess results from repulsion that is amplified near resonances. In the latter case, there should be a deficit of pairs narrow of resonances.

\subsection{Two Important Details}
\label{subsec:remaining}

 Here we discuss two issues plaguing our model.

The first is the presence of near-resonant pairs with $\Delta < 0$ sitting {\it narrow} of MMR. Among the youngest ($<100$\,Myr), and therefore nearly primordial, transiting systems shown in Figure \ref{fig:pp-plane-observed}, these include the outer pair in the three-planet AU Mic system  ($\Delta = -0.013$ from 3:2) and the middle pair in the four-planet V1298 Tau system ($\Delta =-0.027$ from 2:1). 
Additionally, the slightly older ($\sim 600\,\rm{Myr}$) three-planet Kepler-289 system hosts two pairs narrow of the 2:1 MMR with $\Delta =-0.044$ and $-0.047$, for the inner and outer pair, respectively. When we also include the recently-reported TOI-4495 system \citep{Wang2026}, which hosts a pair of near-resonant super-Earths ($\Delta=0.009$ from 2:1), into the sample of young systems in \citet{Dai2024}, the ratio of negative $\Delta$ pairs to positive ones is about one to two.

These narrow-of-resonance orbital arrangements may be hard to accommodate in our ``rattle-and-break" scenario. One possible explanation is that these pairs are repelled from initial $j+1$:$j$ MMRs and stop just short of their current locations near $j$:$j-1$ MMRs. However, this is unlikely both because of the high planetesimal mass required, and because of the fortuitous stalling. Another plausible cause may be competitive repulsion within a chain of pairs. As the simulation in Figure \ref{fig:k223_3br_example} shows, the c-d pair  actually moves mildly closer (not apart) at late times, partly as a result of the strong repulsion experienced by the c-b pair. But it is unclear if such a behavior is universal or if this can drive pairs across  MMRs without being captured. Lastly, these negative $\Delta$ pairs may simply represent the smooth distribution.\footnote{While this could readily explain Kepler-289's architecture, it is more problematic for  AU Mic and V1298 Tau, which both contain additional planet pairs with small positive $\Delta$s.} If so, given that the fraction of near-resonant systems is $\sim 15\%$ overall \citep{Dai2024}, and that the ratio between negative and positive $\Delta$ is one to two, we surmise that no more than 10\% of all systems are largely intact chains, while the rest have undergone instabilities.

Another puzzle is the origin of the rattlers. From where does this percent-level junk, individually wrapped in Pluto-sized bodies,\footnote{A continuous spectrum is allowed as long as the mass is dominated by this scale.} arise? One possibility is initial embryos that fail to grow. Another is debris reformed after incomplete mergers \citep{Wu2024,Li2025}. At the moment, we do not have a ready answer and our theory remains incomplete.

\section{Summary}
\label{sec:summary}
The ``breaking-the-chains" model \citep{Izidoro2017,Izidoro2021,Goldberg2022} has emerged as a  popular theory to explain the architecture of  super-Earths.  This picture suffers from a weakness: the dissipative forces responsible for establishing resonant chain configurations in the first place also leave  the planets in orbital configurations that are exceedingly long-lived or even indefinitely stable. There have to be additional perturbations that rattle them out of such a state.

Borrowing from the insights learned in \citet{Wu2024} for a single pair of planets, we explore how a resonant chain of planets behave under planetesimal scattering.  We show that scattering pushes all pairs within a chain wide of their MMRs. At low scatterer masses, this spreading  initially proceeds along three-body resonances. When the planets are displaced to a few percent wide of two-body MMRs (requiring total perturber mass of order a few percent), the scatterings can kick the system out of the weakening three-body MMRs. When this occurs, the planetary systems are no longer under the protection of MMRs and can become destabilized. Destabilization is more likely in systems with more planets and leaves behind widely-spaced planets on eccentric and inclined orbits. The time to instability appears to depend on the individual perturber mass and can be of order $100$\,Myrs for Pluto-sized bodies. Systems that avoid instability, on the other hand, may be observed as the excess pile-ups outside first-order MMRs, with low eccentricities and mutual inclinations.

 We discuss previous works that rely on much larger perturbers, or that invoke planetesimals injected from much further away. Both, we argue, are unlikely to give rise to the observed distributions. We also point out two major weaknesses of our work: the presence of pairs narrow of resonances, even at  young ages, and the unknown origin of our low-mass perturbers.

\begin{nolinenumbers}
\begin{acknowledgments}
S.H. acknowledges support by the Natural Sciences and Engineering Research Council of Canada (NSERC), funding references CITA 490888-16 and RGPIN-2020-03885  and thanks Renu Malhotra for discussions; YW acknowledges NSERC grant RGPIN-2024-05533.
\end{acknowledgments}
\end{nolinenumbers}
\bibliographystyle{aasjournal}
\bibliography{zotero_lib,more_refs}
\end{document}